\documentclass[aps,twocolumn,amsmath,letterpaper,floatfix]{revtex4}
\usepackage{amssymb}
\usepackage{graphicx}
\usepackage{array}
\usepackage{hhline}
\usepackage{longtable}

\begin{document}

\title{Critical Percolation Phase and Thermal BKT Transition in a Scale-Free Network with Short-Range and Long-Range Random Bonds}
\author{A. Nihat Berker$^{1,2,3}$, Michael Hinczewski,$^{2}$, and Roland R. Netz$^{2}$}
\affiliation{$^1$Faculty of Engineering and Natural Sciences,
Sabanc\i~University, Orhanl\i , Tuzla 34956, Istanbul, Turkey,}
\affiliation{$^2$Department of Physics, Technical University of
Munich, 85748 Garching, Germany,} \affiliation{$^3$Department of
Physics, Massachusetts Institute of Technology, Cambridge,
Massachusetts 02139, U.S.A.}

\begin{abstract}
Percolation in a scale-free hierarchical network is solved exactly
by renormalization-group theory, in terms of the different
probabilities of short-range and long-range bonds.  A phase of
critical percolation, with algebraic
(Berezinskii-Kosterlitz-Thouless) geometric order, occurs in the
phase diagram, in addition to the ordinary (compact) percolating
phase and the non-percolating phase.  It is found that no connection
exists between, on the one hand, the onset of this geometric BKT
behavior and, on the other hand, the onsets of the highly clustered
small-world character of the network and of the thermal BKT
transition of the Ising model on this network.  Nevertheless, both
geometric and thermal BKT behaviors have inverted characters,
occurring where disorder is expected, namely at low bond probability
and high temperature, respectively.  This may be a general property
of long-range networks.


\end{abstract}
\maketitle
\def\s{\rule{0in}{0.28in}}

\begin{figure}
\centering \includegraphics*[scale=0.6]{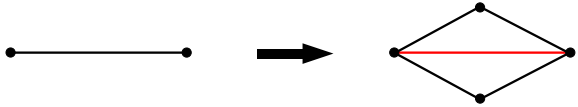}
\caption{(Color online) The scale-free random network is constructed
by the repeated imbedding of the graph as shown in this figure.  The
four edges surrounding the graph here will be imbedded at the next
phase of the construction.  Such surrounding edges of the innermost
graphs of the created infinite network are called the innermost
edges.  Along the innermost edges, a bond occurs with probability q.
Along each of the other edges, a bond occurs with probability p. The
latter are the long-range random bonds. Different realizations are
illustrated in Fig.2.}\label{fig:1}
\end{figure}

\begin{figure}
\centering \includegraphics*[scale=1.2]{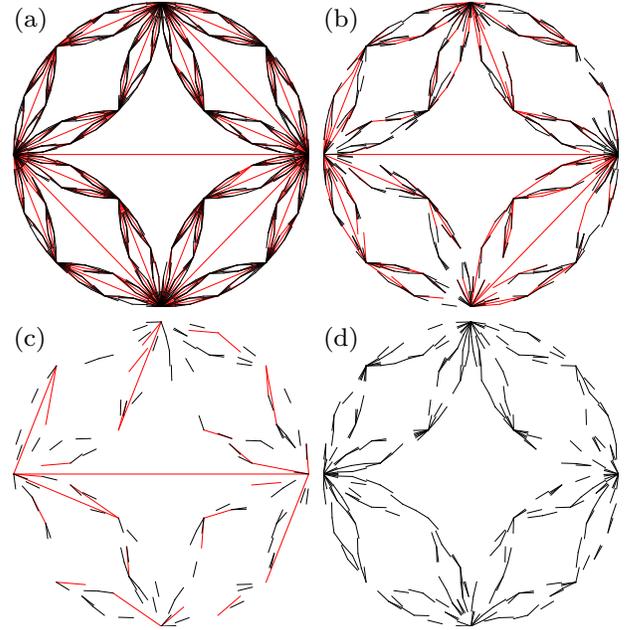}
\caption{(Color online) Different realizations of the random
network: (a) In the compact percolating phase, with $q = p = 0.8$.
(b) In the compact percolating phase, with $q = p = 0.4$. (c) In the
algebraic percolating phase, with $q = p = 0.1$. (d) In the
non-percolating phase, with $q = 0.3, p = 0$.}\label{fig:2}
\end{figure}

\begin{figure}
\includegraphics*[scale=1]{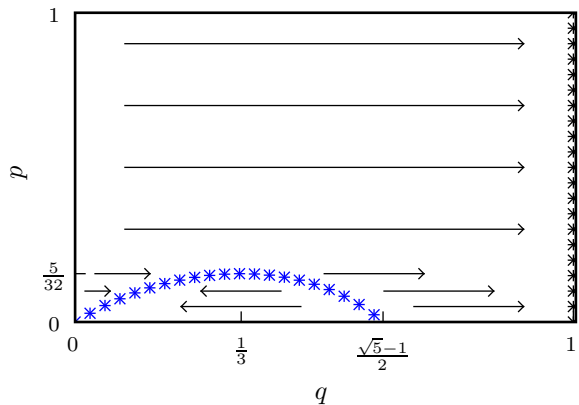}
\caption{(Color online) Renormalization-group flow diagram of
percolation on the network with short-range and long-range random
bonds.}\label{fig:3}
\end{figure}

Scale-free networks are of high current interest
\cite{Albert,Newman,Boccaleti,Dorogovtsev1,Porter}, due to their
ubiquitous occurrence in physical, biological, social, and
information systems and due to their distinctive geometric and
thermal properties.  The geometric properties reflect the
connectivity of the points of the network.  The thermal properties
reflect the interactions, along the geometric lines of connectivity,
between degrees of freedom located at the points of the network.
These interacting degrees of freedom could be voters influencing
each other, persons communicating a disease, etc., and can be
represented by model systems.  Among issues most recently addressed
have been the occurrence of true or algebraic
\cite{Berezinskii,Kosterlitz} order in the geometric or thermal
long-range correlations, and the connection between these geometric
and thermal characteristics. In Ising magnetic systems on a
one-dimensional inhomogeneous lattice
\cite{Costin1,Costin2,Romano,Bundaru} and on an inhomogeneous
growing network \cite{Bauer1}, a Berezinskii-Kosterlitz-Thouless
(BKT) phase in which the thermal correlations between the spins
decay algebraically with distance was found.  In growing networks
\cite{Callaway,Dorogovtsev2,Kim,Lancaster,Bauer2,Coulomb,Bollobas,Krapivsky},
geometric algebraic correlations were seen with the exponential
(non-power-law) scaling of the size of the giant component above the
percolation threshold.  The connection between geometric and thermal
properties was investigated with an Ising magnetic system on a
hierarchical lattice that can be continuously tuned from non-small
world to highly clustered small world via increase of the occurrence
of quenched-random long-range bonds \cite{Hinczewski1}. Whereas in
the non-small-world regime a standard second-order phase transition
was found, when the small-world regime is entered, an inverted BKT
transition was found, with a high-temperature algebraically ordered
phase and a low-temperature phase with true long-range order but
delayed short-range order. Algebraic order in the thermal
correlations has also been found in a community
network.\cite{Hinczewski2} In the current work, the geometric
percolation property of the quenched-random long-range bonds is
studied, aiming to relate the geometric properties to the algebraic
thermal properties. From an exact renormalization-group solution,
surprising results are found both for the geometric properties in
themselves and in their would-be relation to the thermal properties.

The solved infinite network is constructed on a very commonly used
hierarchical lattice \cite{Berker,Kaufman1,Kaufman2} with the
addition of long-range random bonds, as indicated in Fig.~1. The
lattice formed by the innermost edges in the construction explained
in Fig.~1 is indeed one of the most commonly used two-dimensional
hierarchical lattices. In our study, on each of these edges, a bond
occurs with probability $q$. To this hierarchical lattice, all
further-neighbor edges are added between vertices of the same level.
On each of these further-neighbor edges, a bond occurs with
probability $p$, thus completing the random network studied here.
Note that, due to the scale-free nature of this network, phase
transition behaviors as a function of $p$ must be identical along
the lines $q=0$ and $q=p$, which is indeed reflected in the results
below.

Hierarchical lattices provide exact renormalization-group solutions
to network
\cite{Hinczewski1,Hinczewski2,Zhang,Khajeh,Rozenfeld,Wrobel,Boettcher1,Boettcher2,Kaplan}
and other diverse complex problems, as seen in recent works
\cite{Kaufman3,Piolho,Monthus1,Monthus2,Branco,Guven,Ohzeki,Ozcelik,Jorg1,Jorg2,Aral,Gulpinar,Desimoi1,Desimoi2}.
The percolation problem presented by the random network defined
above is also readily solved by renormalization-group theory. The
recursion relation is obtained by replacing graphs at the innermost
level of the random network by equivalent, renormalized,
nearest-neighbor bonds, which thereby occur with renormalized
short-range bond probability
\begin{figure}
\includegraphics*[scale=1]{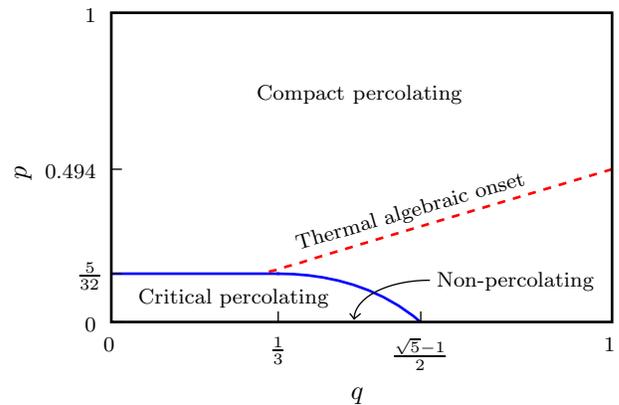}
\caption{(Color online) Geometric phase diagram of the network with
short-range and long-range random bonds, exhibiting compact
percolating, critical percolating, and non-percolating phases. The
dashed line indicates the onset of high-temperature algebraic order
in an Ising magnetic model on this network.  It is thus seen that
this thermal onset has no signature in the geometric
correlations.}\label{fig:4}
\end{figure}
\begin{equation}
\label{eq:1} q' = 1 - (1-q^2)^2(1-p)\,.
\end{equation}
This equation is derived as the probability $1-q'$ of not having any
path across the unit, each $(1-q^2)$ factor being the probability of
one sequence of short-range bonds being missing and $(1-p)$ being
the probability of the long-range bond being missing. The long-range
bond probability $p$ does not get renormalized, similarly to the
thermal long-range interaction in the Ising model lodged on this
network \cite{Hinczewski1}.  The renormalization-group flow of
Eq.\eqref{eq:1} has fixed points at $q=1, p$ arbitrary and at
$q=p=0$.  These fixed points are stable under the
renormalization-group flows and respectively correspond to the sinks
of the ordinary percolating and non-percolating phases. Another
continuum of fixed points is obtained from the solution of
\begin{equation}
\label{eq:2} (1-p)(q^3+q^2-q-1) + 1 = 0\,.
\end{equation}
This equation gives a continuously varying line of fixed points in
the region $0 \leq q \leq (\sqrt{5}-1)/2, 0 \leq p \leq 5/32$.  As
seen in the flow diagram given in Fig.~3, this fixed line starts at
$(q,p) = (0,0)$, continues to $(1/3,5/32)$, and terminates at
$((\sqrt{5}-1)/2, 0)$.  The renormalization-group eigenvalue along
this fixed line is
\begin{equation}
\label{eq:3} \frac{dq'}{dq} = \frac{4q^*}{1+q^*}\,,
\end{equation}
where the fixed point values $q^*$ are determined by $p$ as the
solutions of Eq.(2). Thus, the fixed line is stable in its low-$q$
segment and unstable in its high-$q$ segment. Such reversal of
stability along a fixed line, at $(q,p) = (1/3,5/32)$ here, has also
been seen in the BKT transition of the two-dimensional XY model
\cite{Kosterlitz} and in the Potts critical-tricritical fixed line
in one \cite{Berker1}, two \cite{Nienhuis1}, and three
\cite{Andelman,Nienhuis2} dimensions. The renormalization-group
flows along the entire $q$ direction, at any of the fixed $p$ values
in $0 < p \leq 5/32$, are as seen for the thermal behavior of
antiferromagnetic Potts models \cite{Berker2,Jacobsen1,Jacobsen2}.

As seen in Fig.~3, for $p > 5/32$, renormalization-group flows from
all initial conditions are to the sink $q^* = 1$.  This basin of
attraction is, therefore, an ordinary (compact) percolating
geometric phase. For $p \leq 5/32$, the higher values of $q$ flow to
the sink $q^* = 1$, thereby also being in the ordinary (compact)
percolating geometrical phase. For $0 < p \leq 5/32$, the low values
of $q$ flow to the stable critical fixed point at finite $0 < q^*
\leq 1/3$, thereby being in a critical percolating phase. The
infinite cluster in this phase is not compact at the largest length
scales, but occurs with the bond probability of $q^* $.  For $p =
0$, the low-$q$ phase is the ordinary non-percolating geometric
phase, with sink $q^* = 0$.

The horizontal portion, in Fig.~4, of the phase boundary between the
compact and critical percolating phases is controlled by the fixed
point at $(q,p) =  (1/3,5/32)$ with a marginal direction. The
non-horizontal portion of the phase boundary between the compact and
critical percolating phases is controlled by the unstable fixed line
segment between $(q,p) = (1/3,5/32)$ and $((\sqrt{5}-1)/2, 0)$, and
has continuously varying critical exponents as a function of the
long-range bond probability $p$.  In an interesting contrast,
Kaufman and Kardar \cite{Kaufman4} have found, for percolation on
the Cayley tree with added long-range equivalent-neighbor bonds,
continuously varying critical exponents as a function of the
nearest-neighbor bond probability, between compact percolating and
non-percolating phases. The emergent phase diagram of our current
model is given in Fig.~4.

One of the motivations of our study was to relate the geometric and
thermal properties of this scale-free network.  The Ising magnetic
system located on this network, with Hamiltonian
\begin{equation}\label{eq:4}
-\beta {\cal H} = \sum_{\langle ij \rangle} J s_i s_j\,,
\end{equation}
where $s_i = \pm 1$ at each site $i$, $\langle i j \rangle$
indicates summation over all pairs of sites connected by a
short-range or long-range bond, and the interaction $J>0$ is
ferromagnetic, has an inverted BKT transition, with a
high-temperature algebraically ordered phase, in the compact
percolation phase in the region above the dashed line in Fig.~4.  In
the region below the dashed line in the compact percolation phase,
the Ising transition is an ordinary second-order phase transition.
In the critical percolation phase, the Ising model has no thermal
phase transition.

The rightmost point of the dashed line, $(q,p) = (1,0.494)$, was
calculated in Ref.\cite{Hinczewski1}. It was also seen that this
point separates the non-small-world geometric regime at low $p$ and
the highly clustered, small-world geometric regime at high $p$.  The
flows of $q$ onto $q^* = 1$ given in Eq.\eqref{eq:1} dictate that,
for all $q$ in the currently studied infinite network, highly
clustered small-world behavior occurs for $p \gtrsim 0.494$ and
non-small-world behavior occurs for $p \lesssim 0.494$.

The rest of the dashed line in Fig.~4, with the leftmost point at
$(q,p) = (0.31,5/32)$, has been currently calculated using the
renormalization-group recursion relation for the quenched
probability distribution ${\cal Q}(J)$ for the interactions on the
innermost level,
\begin{multline}
\label{eq:5} {\cal Q}^{(n)}(J'_{i'j'}) = \int \left(
\prod_{ij}^{i'j'}
dJ_{ij} {\cal Q}^{(n-1)}(J_{ij}) \right) dJ_{i'j'}{\cal P}^{(0)}(J_{i'j'})\\
\times \delta(J'_{i'j'} - R(\{J_{ij}\},J_{i'j'}))\,,
\end{multline}
where $(n)$ indicates the distribution after $n$
renormalization-group transformations, ${\cal P}^{(0)}(J)$ is the
initial (double-delta function) and conserved quenched probability
distribution for the interactions on higher levels than innermost,
and $R(\{J_{ij}\},J_{i'j'})$ is the local interaction recursion
relation,
\begin{multline}\label{eq:6}
R(\{J_{ij}\},J_{i'j'}) = \frac{1}{2} \ln \left[
\frac{\cosh(J_{i'k}+J_{kj'})}{\cosh(J_{i'k}-J_{kj'})} \right]\\
+ \frac{1}{2} \ln \left[
\frac{\cosh(J_{i'l}+J_{lj'})}{\cosh(J_{i'l}-J_{lj'})} \right] +
J_{i'j'}\,.
\end{multline}

Thus, it is seen that, although the geometric correlations of this
network show an interesting critical percolating phase, no
quantitative connection exists between the onset of geometric BKT
behavior on the one hand, and the onsets of thermal BKT behavior and
small-world character on the other hand. Qualitatively speaking
however, note that an algebraically ordered geometric phase at low
bond probability is akin to an algebraically ordered thermal phase
at high temperature, both of which are rendered possible on the
network. Thus, inverted algebraic order where disorder is expected
may be a commonly encountered property, both geometrically and
thermally, for long-range random networks.  Finally, we note that
to-date all renormalization-group calculations exploring thermal
behavior on scale-free networks have been done using discrete Ising
or Potts degrees of freedom.  This is because of the compounded
technical burden introduced in position-space renormalization-group
calculations by continuum XY or Heisenberg degrees of freedom, for
example requiring the analysis of the global flows of the order of a
dozen Fourier components of the renormalized
potentials.\cite{Berker2} However, in view of the rich BKT and other
collective phenomena inherent to these continuum-spin models, such
large undertakings may well be worth considering.

\noindent {\em Acknowledgments -} We thank B. Kahng and J.
Kert\'{e}sz for useful conversations.  ANB gratefully acknowledges
the hospitality of the Physics Department of the Technical
University of Munich.  This research was supported by the Alexander
von Humboldt Foundation, the Scientific and Technological Research
Council of Turkey (T\"UB\.ITAK) and by the Academy of Sciences of
Turkey.

\end{document}